\documentclass[english]{article}
\usepackage[T1]{fontenc}
\usepackage[latin9]{inputenc}
\usepackage{geometry}
\usepackage{babel}
\usepackage{amsmath}
\usepackage{graphicx}
\usepackage[unicode=true,pdfusetitle,
 bookmarks=true,bookmarksnumbered=false,bookmarksopen=false,
 breaklinks=false,pdfborder={0 0 1},backref=false,colorlinks=false]
 {hyperref}

\makeatletter
\newcommand{\lyxaddress}[1]{
	\par {\raggedright #1
	\vspace{1.4em}
	\noindent\par}
}

\makeatother

\begin{document}
\title{Exact finite volume expectation values of conserved currents}
\author{Zoltan Bajnok$^{1}$, Istvan Vona$^{1,2}$}
\maketitle

\lyxaddress{\begin{center}
\emph{1. Wigner Research Centre for Physics}\\
\emph{Konkoly-Thege Miklós u. 29-33, 1121 Budapest , Hungary}\\
\emph{and}\\
\emph{2. Roland Eötvös University, }\\
\emph{ Pázmány Péter sétány 1/A, 1117 Budapest, Hungary}
\par\end{center}}
\begin{abstract}
The vacuum expectation values of conserved currents play an essential
role in the generalized hydrodynamics of integrable quantum field
theories. We use analytic continuation to extend these results for
the excited state expectation values in a finite volume. Our formulas
are valid for diagonally scattering theories and incorporate all finite
size corrections. 
\end{abstract}

\section{Introduction}

Recently there have been interesting developments in calculating expectation
values in integrable finite temperature/volume systems. The motivation
came from statistical physics \cite{Castro-Alvaredo:2016cdj} as well
as from the AdS/CFT duality \cite{Bajnok:2014sza}. In the AdS/CFT
duality heavy-heavy light three-point functions can be mapped to expectation
values of local operators in finite volume multiparticle states \cite{Bajnok:2014sza,Hollo:2015cda,Jiang:2015bvm,Jiang:2016dsr}.
In statistical physics the recent developments of the generalized
hydrodynamics require the knowledge of the finite temperature expectation
values of conserved charges and currents as they are the key inputs
in formulating the Euler type hydrodynamic evolution \cite{Castro-Alvaredo:2016cdj,Bertini:2016tmj,Vu:2018iwv}.
There were interesting direct calculations \cite{Borsi:2019tim,Pozsgay:2019xak},
which expressed the current expectation values in spin chain Bethe
states in terms of the charge eigenvalues and the inverse of the Gaudin
matrix. These remarkable compact and simple expressions are also valid
in quantum field theories for finite volume expectation values in
multiparticle states once the exponentially small vacuum polarization
effects are neglected. The aim of our paper is to provide a simple
derivation of this result and to extend it in order to incorporate
all the finite size corrections. As a result we describe exactly the
finite volume excited state expectation values of conserved currents.
In doing so we continue analytically the structural equations of generalized
hydrodynamics \cite{Castro-Alvaredo:2016cdj} and interpret the result
in the finite volume setting.

The paper is organized as follows: In the next section we recall the
results of the generalized hydrodynamics, which can be interpreted
as finite volume vacuum expectation values. We formulate the results
in terms of a pairing between functions, which includes the occupation
number of the quasi-particles as the integration measure. In section
3 we use analytical continuation to modify the pairing to include
also the discrete contributions of physical particles. All formulas
remain the same only the pairing has to be exchanged. Finally we perform
various tests of our results and conclude.

\section{Vacuum expectation values of conserved currents}

In the generalized hydrodynamics of integrable models \cite{Castro-Alvaredo:2016cdj}
conservation laws 
\begin{equation}
\partial_{t}q_{i}(x,t)+\partial_{x}j_{i}(x,t)=0
\end{equation}
play a crucial role. Local thermal equilibrium can be characterized
by temperature like quantities $\beta_{i}$ coupled to the infinite
family of conserved charges, $Q_{i}=\int q_{i}dx$, leading to local
averages
\begin{equation}
\langle\mathcal{O}\rangle=Z^{-1}\text{Tr}(\mathcal{O}e^{-\sum_{j}\beta_{j}Q_{j}})\quad;\qquad Z=\text{Tr}(e^{-\sum_{j}\beta_{j}Q_{j}})
\end{equation}
Here $Q_{1}$ is the energy and $\beta_{1}$ is the inverse of the
temperature (or volume $\beta_{1}=L$ in the finite volume situation).
The collection of the $\beta_{i}$ ``temperatures'' can be traded
for the expectation values of the conserved charges $\langle q_{i}\rangle\propto\partial_{\beta_{i}}\log Z$,
implying that the expectation values of the currents $\langle j_{i}\rangle$
depend on $\langle q_{i}\rangle$ which is the equation of state $\langle j_{i}\rangle=F_{i}(\langle q\rangle)$.
Assuming local thermodynamic equilibrium and that these quantities
vary slowly in space and time they satisfy the continuity equation
$\partial_{t}\langle q_{i}\rangle+\partial_{x}\langle j_{i}\rangle=\partial_{t}\langle q_{i}\rangle+J_{ij}\partial_{x}\langle q_{i}\rangle=0$,
an Euler type hydrodynamic equation. Normal fluid modes diagonalize
$J_{ij}=\partial_{q_{j}}F_{i}$ and propagate as $\partial_{t}n_{i}+v_{i}^{\mathrm{eff}}\partial_{x}n_{i}=0$. 

We focus on a relativistic integrable theory of a single particle
which scatters on itself with the S-matrix $S(\theta_{1}-\theta_{2})$,
where $\theta$ is the rapidity which parametrizes the energy and
momentum as $E(\theta)=m\cosh\theta$, $p(\theta)=m\sinh\theta$.
In thermal equilibrium the expectation values of charges can be calculated
from the densities of quasi-particles $\rho(\theta)$ and the charge
eigenvalue on a one-particle state $h_{i}(\theta)$\footnote{Here $h_{1}(\theta)=m\cosh\theta,$ and there are higher spin conserved
charges $h_{2n-1}(\theta)\propto\cosh(n\theta)$, $h_{2n}(\theta)\propto\sinh(n\theta)$
for infinitely many odd $n$s.} as 
\begin{equation}
\langle q_{i}\rangle=\int\frac{d\theta}{2\pi}\rho(\theta)h_{i}(\theta)\label{eq:charge}
\end{equation}
Here and from now on all integrals go from $-\infty$ to $\infty$.
Thus the state in a thermal equilibrium can be represented either
by $\beta_{i}$ or by $q_{i}$ or alternatively by $\rho$. The normal
modes, however are neither of these, instead they are related to the
occupation number $n$:
\begin{equation}
n(\theta)=\frac{1}{1+e^{\epsilon(\theta)}}
\end{equation}
which can be calculated from the Thermodynamic Bethe ansatz (TBA)
equation \cite{Zamolodchikov:1989cf}
\begin{equation}
\epsilon(\theta)=\sum_{i}\beta_{i}h_{i}(\theta)-\int\frac{du}{2\pi}\varphi(\theta-u)\log(1+e^{-\epsilon(u)})
\end{equation}
where $\varphi(\theta)=-i\partial_{\theta}\log S(\theta)$. For later
generalizations we introduce a pairing including the occupation number
as
\begin{equation}
g(\theta)\circ h(\theta)=\int g(\theta)h(\theta)\frac{n(\theta)d\theta}{2\pi}
\end{equation}
The TBA equation after integration by parts takes the form 
\begin{equation}
\epsilon(\theta)=\sum_{i}\beta_{i}h_{i}(\theta)-i\log S(\theta-u)\circ\partial_{u}\epsilon(u)
\end{equation}
It is also useful to introduce dressed quantities which satisfy
\begin{equation}
g^{\mathrm{dr}}(\theta)=g(\theta)+\varphi(\theta-u)\circ g^{\mathrm{dr}}(u)
\end{equation}
since then the particle density can be written in terms of the occupation
number as
\begin{equation}
\rho(\theta)=n(\theta)(p')^{\mathrm{dr}}(\theta)
\end{equation}
where $p'(\theta)=dp(\theta)/d\theta$. This leads to the charge expectation
value 
\begin{equation}
\langle q_{i}\rangle=(p')^{\mathrm{dr}}(\theta)\circ h_{i}(\theta)=p'(\theta)\circ h_{i}^{\mathrm{dr}}(\theta)\label{eq:qiev}
\end{equation}
In the second equality we used the fact that the dressing operator
$(1-\varphi\circ)^{-1}$ is symmetric wrt. the pairing. From relativistic
invariance it follows \cite{Castro-Alvaredo:2016cdj} that the current
expectation values take the form
\begin{equation}
\langle j_{i}\rangle=E'(\theta)\circ h_{i}^{\mathrm{dr}}(\theta)=(E')^{\mathrm{dr}}(\theta)\circ h_{i}(\theta)\label{eq:jiev}
\end{equation}
Comparing $\langle j_{i}\rangle$ to $\langle q_{i}\rangle$ we can
extract the effective velocity of the quasi-particles $v^{\mathrm{eff}}(\theta)=(E')^{\mathrm{dr}}/(p')^{\mathrm{dr}}$.

These results can also be obtained from the Leclair-Mussardo (LM)
formula \cite{Leclair:1999ys}
\begin{equation}
\langle\mathcal{O}\rangle=\sum_{n=0}^{\infty}\frac{1}{n!}\prod_{k=1}^{n}\int\frac{n(\theta_{k})d\theta_{k}}{2\pi}F_{2n,c}^{{\cal O}}(\theta_{1},\dots,\theta_{n})\label{eq:LMthermal}
\end{equation}
by taking into account that the connected form factors of the conserved
charges and currents are 
\begin{equation}
F_{2n,c}^{q_{i}}(\theta_{1},\dots,\theta_{n})=E(\theta_{1})\varphi(\theta_{1}-\theta_{2})\dots\varphi(\theta_{n-1}-\theta_{n})h_{i}(\theta_{n})+\text{permutations}
\end{equation}
\begin{equation}
F_{2n,c}^{j_{i}}(\theta_{1},\dots,\theta_{n})=p(\theta_{1})\varphi(\theta_{1}-\theta_{2})\dots\varphi(\theta_{n-1}-\theta_{n})h_{i}(\theta_{n})+\text{permutations}
\end{equation}
Indeed, expanding the dressing operator $(1-\varphi\circ)^{-1}$ in
(\ref{eq:qiev},\ref{eq:jiev}) leads to the LM formula.

Using the fact that 
\begin{equation}
\partial_{\beta_{i}}\epsilon(\theta)=h_{i}^{\mathrm{dr}}(\theta)
\end{equation}
we can express the charge and current expectation values as 
\begin{equation}
\langle q_{i}\rangle=-\partial_{\beta_{i}}\int\frac{dp(\theta)}{2\pi}\log(1+e^{-\epsilon(\theta)})=-\partial_{\beta_{i}}(p(\theta)\circ\partial_{\theta}\epsilon(\theta))\label{eq:qbeta}
\end{equation}
\begin{equation}
\langle j_{i}\rangle=-\partial_{\beta_{i}}\int\frac{dE(\theta)}{2\pi}\log(1+e^{-\epsilon(\theta)})=-\partial_{\beta_{i}}(E(\theta)\circ\partial_{\theta}\epsilon(\theta))\label{eq:jbeta}
\end{equation}
These expectation values are valid in a local thermal equilibrium
specified by the ``temperatures'' $\beta_{i}$.

To make contact with the finite volume description in the crossed
channel we need to choose $\beta_{1}=L$ to be the volume and put
all other $\beta_{i}$ to zero\footnote{Keeping $\beta_{i}$ nonzero would imply twisted boundary conditions
by the conserved charges \cite{Ahn:2011xq,Hernandez-Chifflet:2019sua} } . Thus the TBA equation is understood as the generating function
of the expectation values of conserved quantities where, after differentiation
in (\ref{eq:qbeta},\ref{eq:jbeta}), we have to take $\beta_{i}=\delta_{1i}L$.
In this simplified situation $\partial_{\theta}\epsilon(\theta)=L(E')^{\mathrm{dr}}(\theta)$
and we can simplify the current expectation values as
\begin{equation}
\langle j_{i}\rangle=\frac{1}{L}h_{i}(\theta)\circ\partial_{\theta}\epsilon(\theta)=\frac{1}{L}\int\frac{d\theta}{2\pi}h_{i}'(\theta)\log(1+e^{-\epsilon(\theta)})
\end{equation}
but the same is not true for the charges. From the relativistic invariance
we can reformulate the finite temperature partition function and averages
in the mirror channel. In the Euclidean version it is obtained by
a $\frac{\pi}{2}$ rotation. This is an imaginary Lorentz transformation
with rapidity $i\frac{\pi}{2}$: $\theta\to\theta^{\gamma}=\theta+\frac{i\pi}{2}$,
for which coordinates transform as $(x,t)\to(it,ix)$, while currents
and charges as $(j,q)\to(iq,ij)$, in particular $(p,E)\to(iE,ip)$.
This transformation squares to the crossing transformation, which
acts as $(j,q)\to-(j,q)$ and changes particles to antiparticles.
In the finite volume channel, indicated by a subscript $L$, the LM
formula takes the form 
\begin{equation}
\langle0\vert\mathcal{O}\vert0\rangle_{L}=\sum_{n=0}^{\infty}\frac{1}{n!}\prod_{k=1}^{n}\int\frac{n(\theta_{k})d\theta_{k}}{2\pi}F_{2n,c}^{{\cal O}}(\theta_{1}^{\gamma},\dots,\theta_{n}^{\gamma})\label{eq:LMvolume}
\end{equation}
 which implies 
\begin{equation}
\langle0\vert q_{k}\vert0\rangle_{L}=i(E')^{\mathrm{dr}}(\theta)\circ h_{k}(\theta^{\gamma})=i\langle j_{k}^{\gamma}\rangle
\end{equation}
where by $j_{k}^{\gamma}$ we mean that  we use $h_{k}^{\gamma}(\theta)=h_{k}(\theta^{\gamma})$
for the corresponding charge eigenvalue. In particular, the finite
volume vacuum expectation value of the conserved charges is 
\begin{equation}
\langle0\vert q_{k}\vert0\rangle_{L}=\frac{i}{L}\int\frac{d\theta}{2\pi}h_{k}'(\theta+\frac{i\pi}{2})\log(1+e^{-\epsilon(\theta)})
\end{equation}
Evaluating this expression for the energy, $h_{1}(\theta)=m\cosh\theta$,
gives 
\begin{equation}
L\langle0\vert q_{1}\vert0\rangle_{L}=-m\int\frac{d\theta}{2\pi}\cosh\theta\,\log(1+e^{-\epsilon(\theta)})
\end{equation}
which agrees with the groundstate energy $E_{0}(L)$ coming from the
saddle point value of the partition function \cite{Zamolodchikov:1989cf}.

Similarly, the finite volume vacuum expectation value of the currents
can be expressed as
\begin{equation}
\langle0\vert j_{i}\vert0\rangle_{L}=i(p')^{\mathrm{dr}}(\theta)\circ h_{i}(\theta^{\gamma})=i\langle q_{i}^{\gamma}\rangle
\end{equation}
In the following we generalize these results for finite volume excited
states.

\section{Excited state expectation values of conserved currents}

It was observed in \cite{Dorey:1996re} that excited state TBA equations
can be obtained from the ground-state one by analytical continuations.
The idea is that by doing an analytical continuation in the volume/temperature
to complex values a pole singularity of $n(\theta)$ might cross the
real integration contour whose residue should be picked up and added
as a source term even when the volume is continued back to its physical
value. The resulting TBA equation describes excited multiparticle
states in the finite volume channel. 

In the thermal channel the situation might be interpreted as the presence
of some defect lines which correspond to physical particles propagating
in the crossed channel. These defects then modify the thermal equilibrium
and change the quasi-particle density \cite{Bajnok:2004jd}. As a
result we need to use the new densities and occupation numbers to
calculate averages in this situation, which we denote by the same
symbol as before. 

\begin{figure}
\begin{centering}
\includegraphics[width=8cm]{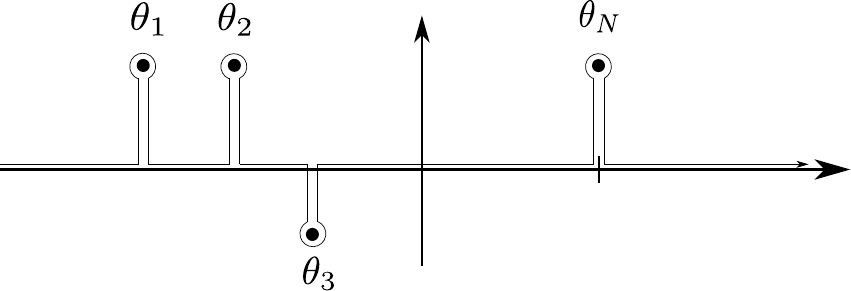}
\par\end{centering}
\caption{Schematical integration contour for excited states. In the sinh-Gordon
model the singularities have imaginary part $i\frac{\pi}{2}$. In
the scaling Lee-Yang model they are symmetric for the real line.}

\label{fig:integration}
\end{figure}

In analyzing the finite volume excited state expectation values in
the sinh-Gordon model \cite{Bajnok:2019yik} it turned out that all
effects coming from the analytical continuation can be encoded into
the pairing. Thus we expect that all formulae remain the same as the
groundstate ones except that the pairing has to be replaced with a
new pairing:

\begin{equation}
g(\theta)\bullet h(\theta)=\sum_{j}\frac{\eta_{j}if(\theta_{j})g(\theta_{j})}{\partial_{\theta}\epsilon(\theta)\vert_{\theta_{j}}}+\int g(\theta)h(\theta)\frac{n(\theta)d\theta}{2\pi}
\end{equation}
Formally we can represent the effect of the continuation with a modified
contour as shown on Figure \ref{fig:integration}. The residues $\eta_{j}$
are $1$ for poles on the upper and $-1$ for the lower half-plane.

The rapidities $\theta_{i}$ are determined by $\epsilon(\theta_{i})=i\pi(2m_{i}+1)$
and we have to take $\partial_{\theta}\epsilon(\theta)\vert_{\theta_{j}}=\sum_{i}\beta_{i}h_{i}'(\theta_{j})+\varphi(\theta_{j}-u)\bullet\partial_{u}\epsilon(u)$.
In the modified convolution the occupation number is $n=1/(1+e^{\epsilon})$,
where now $\epsilon$ satisfies the excited state TBA equation, which
again can be obtained via the new convolution:
\begin{align}
\epsilon(\theta) & =\sum_{k}\beta_{k}h_{k}(\theta)-i\log S(\theta-u)\bullet\partial_{u}\epsilon(u)\\
 & =\sum_{k}\beta_{k}h_{k}(\theta)+\sum_{k}\eta_{k}\log S(\theta-\theta_{k})-i\log S(\theta-u)\circ\partial_{u}\epsilon(u)\nonumber 
\end{align}
The excited state expectation values of the conserved charges and
currents are simply 
\begin{equation}
\langle q_{i}\rangle=(p')^{\mathbf{dr}}(\theta)\bullet h_{i}(\theta)=p'(\theta)\bullet h_{i}^{\mathbf{dr}}(\theta)\label{eq:qiddr}
\end{equation}
\begin{equation}
\langle j_{i}\rangle=E'(\theta)\bullet h_{i}^{\mathrm{\mathbf{dr}}}(\theta)=(E')^{\mathbf{dr}}(\theta)\bullet h_{i}(\theta)\label{eq:jiddr}
\end{equation}
where the dressed quantities, indicated by boldface, are obtained
by means of the new convolution 
\begin{align}
g^{\mathbf{dr}}(\theta) & =g(\theta)+\varphi(\theta-u)\bullet g^{\mathbf{dr}}(u)
\end{align}
Even simpler expressions can be obtained for the expectation values
as
\begin{equation}
\langle q_{i}\rangle=-\partial_{\beta_{i}}(p(\theta)\bullet\partial_{\theta}\epsilon(\theta))
\end{equation}
\begin{equation}
\langle j_{i}\rangle=-\partial_{\beta_{i}}(E(\theta)\bullet\partial_{\theta}\epsilon(\theta))
\end{equation}
These are the main results of the paper. Since they were not really
derived, merely conjectured based on previous experiences, we perform
various consistency checks and elaborate the details. 

As a start we separate the contribution of the physical particles
and the quasi-particles. In doing so we rewrite the new convolution
in terms of the old one. 

\begin{align}
\langle j_{i}\rangle & =-\partial_{\beta_{i}}(i\sum_{k}\eta_{k}E(\theta_{k})+E(\theta)\bullet\partial_{\theta}\epsilon(\theta))\\
 & =-i\sum_{k}\eta_{k}E'(\theta_{k})\partial_{\beta_{i}}\theta_{k}-\partial_{\beta_{i}}(E(\theta)\circ\partial_{\theta}\epsilon(\theta))\nonumber 
\end{align}
In order to calculate $\partial_{\beta_{i}}\theta_{k}$ we take the
quantization condition $\epsilon(\theta_{k})=i\pi(2m_{k}+1)$:
\begin{equation}
\epsilon(\theta_{k})=\sum_{i}\beta_{i}h_{i}(\theta_{k})+\sum_{j}\eta_{j}\log S(\theta_{k}-\theta_{j})-i\log S(\theta_{k}-u)\circ\partial_{u}\epsilon(u)
\end{equation}
and differentiate wrt. $\beta_{i}$. We can do it in two different
ways. 

In the first we differentiate $\epsilon(\theta)$ by keeping $\theta_{k}$
independent of $\beta_{i}$ and then we take into account the $\beta_{i}$
dependence of all $\theta_{j}$s: 
\begin{equation}
0=\partial_{\beta_{i}}\epsilon(\theta_{k})=h_{i}^{\mathrm{dr}}(\theta_{k})+\sum_{j}\partial_{\theta_{j}}\epsilon(\theta_{k})\partial_{\beta_{i}}\theta_{j}
\end{equation}
In the second term we recognize the Gaudin matrix 
\begin{equation}
G_{jk}=-i\partial_{\theta_{j}}\epsilon(\theta_{k})=-i(\delta_{jk}\partial_{\theta}\epsilon(\theta)+\partial_{\theta_{j}}\epsilon(\theta))\vert_{\theta_{k}}=\delta_{jk}D^{\mathrm{dr}}(\theta_{k})-\varphi_{j}^{\mathrm{dr}}(\theta_{k})
\end{equation}
where 
\begin{equation}
D(\theta)=-i\sum_{l}\beta_{l}h_{l}'(\theta)+\sum_{j}\varphi_{j}(\theta)\quad;\qquad\varphi_{j}(\theta)=\eta_{j}\varphi(\theta-\theta_{j})
\end{equation}

Alternatively, we can separate the $\beta_{i}$-dependence of $\theta_{k}$
in the argument: 
\begin{equation}
0=\partial_{\beta_{i}}\epsilon(\theta_{k})=\partial_{\theta}\epsilon(\theta)\vert_{\theta_{k}}\frac{\partial\theta_{k}}{\partial\beta_{i}}+\partial_{\beta_{i}}\epsilon(\theta)\vert_{\theta_{k}}
\end{equation}
This formula can be used to show that $\partial_{\beta_{i}}\epsilon(\theta)=h_{i}^{\mathbf{dr}}(\theta)$.
By putting together these contributions we obtain 
\begin{equation}
\langle j_{i}\rangle=\sum_{k,l}\eta_{k}E'(\theta_{k})G_{kl}^{-1}h_{i}^{\mathrm{dr}}(\theta_{l})+E'(\theta)\circ h_{i}^{\mathbf{dr}}(\theta)
\end{equation}
In order to have a complete separation into physical and quasi-particles
we rewrite the new dressing in terms of the old one. In doing so we
note that 
\begin{align}
g^{\mathbf{dr}}(\theta) & =g(\theta)+\sum_{j}\frac{\varphi_{j}(\theta)}{D^{\mathrm{dr}}(\theta_{j})}g(\theta_{j})+\varphi(\theta-u)\circ g^{\mathbf{dr}}(u)=\sum_{n=0}^{\infty}[\varphi(\theta-u)\bullet]^{n}g(u)
\end{align}
In each term we can use either the discrete or the continuous parts
of the convolution. The continuous part dresses up $g(\theta)$ and
$\varphi_{j}(\theta)$ leading to 
\begin{equation}
g^{\mathbf{dr}}(\theta)=g^{\mathrm{dr}}(\theta)+\sum_{j}\varphi_{j}^{\mathrm{dr}}(\theta)G_{jk}^{-1}g^{\mathrm{dr}}(\theta_{k})
\end{equation}
Here we used that $G_{jk}=D^{\mathrm{dr}}(\theta_{k})(\delta_{jk}-\varphi_{j}^{\mathrm{dr}}(\theta_{k})/D^{\mathrm{dr}}(\theta_{k}))$
in recognizing its inverse. Plugging this back into the current expectation
value we obtain a form involving only the old dressing and convolutions:
\begin{equation}
\langle j_{i}\rangle=\sum_{k,l}\eta_{k}(E')^{\mathrm{dr}}(\theta_{k})G_{kl}^{-1}h_{i}^{\mathrm{dr}}(\theta_{l})+E'(\theta)\circ h_{i}^{\mathrm{dr}}(\theta)
\end{equation}
We note that one can show in general that $f(\theta)\bullet g^{\mathbf{dr}}(\theta)=\sum_{k,l}f^{\mathrm{dr}}(\theta_{k})G_{kl}^{-1}g^{\mathrm{dr}}(\theta_{l})+f(\theta)\circ g^{\mathrm{dr}}(\theta)$.
Writing $h_{i}^{\mathrm{dr}}(\theta_{l})=h_{i}(\theta_{l})+\varphi(\theta_{l}-u)\circ h_{i}^{\mathrm{dr}}(u)$
we can observe that the leading part of the result, i.e. the term
without any integration, is the same which was obtained in \cite{Borsi:2019tim,Pozsgay:2019xak}
in a more complicated way. 

An analogous calculation results in the charge expectation value
\begin{equation}
\langle q_{i}\rangle=\sum_{k,l}\eta_{k}(p')^{\mathrm{dr}}(\theta_{k})G_{kl}^{-1}h_{i}^{\mathrm{dr}}(\theta_{l})+p'(\theta)\circ h_{i}^{\mathrm{dr}}(\theta)
\end{equation}

In the following we use these results to calculate the finite volume
excited state expectation values. For this reason we again take $\beta_{i}=L\delta_{i,1}$
and use the relation 
\begin{equation}
\partial_{\theta}\epsilon(\theta)=LE'(\theta)+i\sum_{k}\eta_{k}\varphi(\theta-\theta_{k})+\varphi(\theta-u)\circ\partial_{u}\epsilon(u)=L(E')^{\mathbf{dr}}(\theta)
\end{equation}
to obtain
\begin{equation}
\langle j_{k}\rangle=\frac{1}{L}\partial_{\theta}\epsilon(\theta)\bullet h_{k}(\theta)=\frac{1}{L}(i\sum_{j}\eta_{j}h_{k}(\theta_{j})+\int\frac{d\theta}{2\pi}h_{k}'(\theta)\log(1+e^{-\epsilon(\theta)}))
\end{equation}

In the finite volume interpretation the expectation values correspond
to excited states diagonal matrix elements

\begin{equation}
\langle\{\theta\}\vert q_{k}\vert\{\theta\}\rangle_{L}=i(E')^{\mathrm{dr}}(\theta)\bullet h_{k}(\theta^{\gamma})=i\langle j_{k}^{\gamma}\rangle\label{eq:qev}
\end{equation}

\begin{equation}
\langle\{\theta\}\vert j_{k}\vert\{\theta\}\rangle_{L}=i(p')^{\mathrm{dr}}(\theta)\bullet h_{k}(\theta^{\gamma})=i\langle q_{k}^{\gamma}\rangle\label{eq:jev}
\end{equation}
where $\{\theta\}\equiv\{\theta_{1},\dots,\theta_{n}\}$ represents
the excited state. The parameters $\theta_{i}$ appearing in the formulas
above are not the rapidities of the particles, but they are related
to them, although in a model-dependent way. 

In the following we elaborate further on these results. For the expectation
value of the conserved charge we can write 
\begin{equation}
L\langle\{\theta\}\vert q_{k}\vert\{\theta\}\rangle_{L}=-\sum_{j}\eta_{j}h_{k}(\theta_{j}+\frac{i\pi}{2})+i\int\frac{d\theta}{2\pi}h_{k}'(\theta+\frac{i\pi}{2})\log(1+e^{-\epsilon(\theta)})
\end{equation}
In the sinh-Gordon model $\eta_{k}=1$ and $\theta_{k}=\bar{\theta}_{k}+\frac{i\pi}{2}$,
where $\bar{\theta}_{k}$ is the rapidity of the particle, thus our
formula reproduces the charge eigenvalue correctly, which asymptotically
takes the form $Q_{i}=\sum_{k}h_{i}(\bar{\theta}_{k})$. 

For the current expectation value we have no such a simplification:
\begin{equation}
\langle\{\theta\}\vert j_{k}\vert\{\theta\}\rangle_{L}=i\sum_{m,l}\eta_{m}(p')^{\mathrm{dr}}(\theta_{m})G_{ml}^{-1}h_{k}(\theta_{l}+\frac{i\pi}{2})^{\mathrm{dr}}+ip'(\theta)\circ h_{k}(\theta+\frac{i\pi}{2})^{\mathrm{dr}}
\end{equation}
where in the last terms $h_{k}(\theta+\frac{i\pi}{2})^{\mathrm{dr}}$
means that $h_{k}(\theta+\frac{i\pi}{2})$ is dressed. This formula
is the main result of our paper, which describes the exact finite
volume expectation value of conserved currents. It is equivalent to
(\ref{eq:jev}) but written in the form where the polynomial and exponential
finite size corrections are separated. Indeed, since the convolution
kernel $n$ is exponentially small we can forget the dressing operator
in each term to obtain the asymptotic results, which in the sinh-Gordon
case, reads as
\begin{equation}
\langle\{\theta\}\vert j_{i}\vert\{\theta\}\rangle_{L}=\sum_{k,l}E'(\bar{\theta}_{k})\bar{G}_{kl}^{-1}h_{i}(\bar{\theta}_{l})
\end{equation}
Recall that the Gaudin matrix was also the dressed version of its
asymptotic form $\bar{G}_{jk}=\delta_{jk}D(\theta_{k})-\varphi_{j}(\theta_{k})$.
This formula agrees with the recent direct calculations in \cite{Borsi:2019tim,Pozsgay:2019xak}.
We also checked these formulas in the sinh-Gordon theory  against
the generalization of the LM formula for excited states \cite{Pozsgay:2013jua}.
In doing so we had to take into account that \cite{Pozsgay:2013jua}
is valid in the thermal channel for operators with spins. In the finite
volume channel the quasi-particle arguments of the connected form
factors should be shifted, similarly how (\ref{eq:LMvolume}) is shifted
compared to (\ref{eq:LMthermal}), while the discrete rapidities take
their physical values. 

Let us finally point out that in deriving our result we used the analytical
continuation of the charge eigenvalue (\ref{eq:charge}) in the thermal
channel and not the current eigenvalue. 

\section{Conclusions}

Using the analytical continuation method for the vacuum expectation
values of conserved charges and currents we managed to derive exact
excited state expectation values. We performed this calculation both
in the thermal and finite volume settings, where the role of the currents
and charges are exchanged. In the finite volume situation the charges
act diagonally and have simple eigenvalues, while currents act nondiagonally
and have more complicated expectation values. In the asymptotic limit,
when vacuum polarization effects are neglected the currents expectation
values can be expressed in terms of the charge eigenvalues and the
inverse of the Gaudin matrix in agreement with previous calculations
\cite{Borsi:2019tim,Pozsgay:2019xak}. Our results provide all the
finite size corrections to the asymptotical formulas valid in a diagonally
scattering integrable theory with a single species. Multiparticle
generalizations for diagonal scatterings are straightforward as well
as the extension for flows generated by other conserved charges. It
would be very nice to derive similar formulas for non-diagonally scattering
theories. The simplest of such results was obtained for the topological
current in the sine-Gordon theory in \cite{Hegedus:2017muz}. 

\subsection*{Acknowledgments}

We thank Balázs Pozsgay, Márton Kormos and Gábor Takács for the useful
discussions and comments and the NKFIH research Grant K116505 for
support. 

\bibliographystyle{utphys}
\bibliography{j1ev}

\providecommand{\href}[2]{#2}\begingroup\raggedright\begin{thebibliography}{10}

\bibitem{Castro-Alvaredo:2016cdj}
O.~A. Castro-Alvaredo, B.~Doyon, and T.~Yoshimura, ``{Emergent hydrodynamics in
  integrable quantum systems out of equilibrium},''
  \href{http://dx.doi.org/10.1103/PhysRevX.6.041065}{{\em Phys. Rev.}
  {\bfseries X6} no.~4, (2016) 041065},
\href{http://arxiv.org/abs/1605.07331}{{\ttfamily arXiv:1605.07331
  [cond-mat.stat-mech]}}.

\bibitem{Bajnok:2014sza}
Z.~Bajnok, R.~A. Janik, and A.~Wereszczynski, ``{HHL correlators, orbit
  averaging and form factors},''
  \href{http://dx.doi.org/10.1007/JHEP09(2014)050}{{\em JHEP} {\bfseries 09}
  (2014) 050},
\href{http://arxiv.org/abs/1404.4556}{{\ttfamily arXiv:1404.4556 [hep-th]}}.

\bibitem{Hollo:2015cda}
L.~Hollo, Y.~Jiang, and A.~Petrovskii, ``{Diagonal Form Factors and
  Heavy-Heavy-Light Three-Point Functions at Weak Coupling},''
  \href{http://dx.doi.org/10.1007/JHEP09(2015)125}{{\em JHEP} {\bfseries 09}
  (2015) 125},
\href{http://arxiv.org/abs/1504.07133}{{\ttfamily arXiv:1504.07133 [hep-th]}}.

\bibitem{Jiang:2015bvm}
Y.~Jiang and A.~Petrovskii, ``{Diagonal form factors and hexagon form
  factors},'' \href{http://dx.doi.org/10.1007/JHEP07(2016)120}{{\em JHEP}
  {\bfseries 07} (2016) 120},
\href{http://arxiv.org/abs/1511.06199}{{\ttfamily arXiv:1511.06199 [hep-th]}}.

\bibitem{Jiang:2016dsr}
Y.~Jiang, ``{Diagonal Form Factors and Hexagon Form Factors II. Non-BPS Light
  Operator},'' \href{http://dx.doi.org/10.1007/JHEP01(2017)021}{{\em JHEP}
  {\bfseries 01} (2017) 021},
\href{http://arxiv.org/abs/1601.06926}{{\ttfamily arXiv:1601.06926 [hep-th]}}.

\bibitem{Bertini:2016tmj}
B.~Bertini, M.~Collura, J.~De~Nardis, and M.~Fagotti, ``{Transport in
  Out-of-Equilibrium $XXZ$ Chains: Exact Profiles of Charges and Currents},''
  \href{http://dx.doi.org/10.1103/PhysRevLett.117.207201}{{\em Phys. Rev.
  Lett.} {\bfseries 117} no.~20, (2016) 207201},
\href{http://arxiv.org/abs/1605.09790}{{\ttfamily arXiv:1605.09790
  [cond-mat.stat-mech]}}.

\bibitem{Vu:2018iwv}
D.-L. Vu and T.~Yoshimura, ``{Equations of state in generalized
  hydrodynamics},'' \href{http://dx.doi.org/10.21468/SciPostPhys.6.2.023}{{\em
  SciPost Phys.} {\bfseries 6} no.~2, (2019) 023},
\href{http://arxiv.org/abs/1809.03197}{{\ttfamily arXiv:1809.03197
  [cond-mat.stat-mech]}}.

\bibitem{Borsi:2019tim}
M.~Borsi, B.~Pozsgay, and L.~Pristyak, ``{Current operators in Bethe Ansatz and
  Generalized Hydrodynamics: An exact quantum/classical correspondence},''
\href{http://arxiv.org/abs/1908.07320}{{\ttfamily arXiv:1908.07320
  [cond-mat.stat-mech]}}.

\bibitem{Pozsgay:2019xak}
B.~Pozsgay, ``{Current operators in integrable spin chains: lessons from long
  range deformations},''
\href{http://arxiv.org/abs/1910.12833}{{\ttfamily arXiv:1910.12833
  [cond-mat.stat-mech]}}.

\bibitem{Zamolodchikov:1989cf}
A.~B. Zamolodchikov, ``{Thermodynamic Bethe Ansatz in Relativistic Models.
  Scaling Three State Potts and Lee-Yang Models},''
\href{http://dx.doi.org/10.1016/0550-3213(90)90333-9}{{\em Nucl. Phys.}
  {\bfseries B342} (1990) 695--720}.

\bibitem{Leclair:1999ys}
A.~Leclair and G.~Mussardo, ``{Finite temperature correlation functions in
  integrable QFT},''
  \href{http://dx.doi.org/10.1016/S0550-3213(99)00280-1}{{\em Nucl. Phys.}
  {\bfseries B552} (1999) 624--642},
\href{http://arxiv.org/abs/hep-th/9902075}{{\ttfamily arXiv:hep-th/9902075
  [hep-th]}}.

\bibitem{Ahn:2011xq}
C.~Ahn, Z.~Bajnok, D.~Bombardelli, and R.~I. Nepomechie, ``{TBA, NLO Luscher
  correction, and double wrapping in twisted AdS/CFT},''
  \href{http://dx.doi.org/10.1007/JHEP12(2011)059}{{\em JHEP} {\bfseries 12}
  (2011) 059},
\href{http://arxiv.org/abs/1108.4914}{{\ttfamily arXiv:1108.4914 [hep-th]}}.

\bibitem{Hernandez-Chifflet:2019sua}
G.~Hernandez-Chifflet, S.~Negro, and A.~Sfondrini, ``{Flow equations for
  generalised $T\bar{T}$ deformations},''
\href{http://arxiv.org/abs/1911.12233}{{\ttfamily arXiv:1911.12233 [hep-th]}}.

\bibitem{Dorey:1996re}
P.~Dorey and R.~Tateo, ``{Excited states by analytic continuation of TBA
  equations},'' \href{http://dx.doi.org/10.1016/S0550-3213(96)00516-0}{{\em
  Nucl. Phys.} {\bfseries B482} (1996) 639--659},
\href{http://arxiv.org/abs/hep-th/9607167}{{\ttfamily arXiv:hep-th/9607167
  [hep-th]}}.

\bibitem{Bajnok:2004jd}
Z.~Bajnok and A.~George, ``{From defects to boundaries},''
  \href{http://dx.doi.org/10.1142/S0217751X06025262}{{\em Int. J. Mod. Phys.}
  {\bfseries A21} (2006) 1063--1078},
\href{http://arxiv.org/abs/hep-th/0404199}{{\ttfamily arXiv:hep-th/0404199
  [hep-th]}}.

\bibitem{Bajnok:2019yik}
Z.~Bajnok and F.~Smirnov, ``{Diagonal finite volume matrix elements in the
  sinh-Gordon model},''
  \href{http://dx.doi.org/10.1016/j.nuclphysb.2019.114664}{{\em Nucl. Phys.}
  {\bfseries B945} (2019) 114664},
\href{http://arxiv.org/abs/1903.06990}{{\ttfamily arXiv:1903.06990 [hep-th]}}.

\bibitem{Pozsgay:2013jua}
B.~Pozsgay, ``{Form factor approach to diagonal finite volume matrix elements
  in Integrable QFT},'' \href{http://dx.doi.org/10.1007/JHEP07(2013)157}{{\em
  JHEP} {\bfseries 07} (2013) 157},
\href{http://arxiv.org/abs/1305.3373}{{\ttfamily arXiv:1305.3373 [hep-th]}}.

\bibitem{Hegedus:2017muz}
A.~Hegedus, ``{Lattice approach to finite volume form-factors of the Massive
  Thirring (Sine-Gordon) model},''
  \href{http://dx.doi.org/10.1007/JHEP08(2017)059}{{\em JHEP} {\bfseries 08}
  (2017) 059},
\href{http://arxiv.org/abs/1705.00319}{{\ttfamily arXiv:1705.00319 [hep-th]}}.

\end{thebibliography}\endgroup

\end{document}